\documentclass[acus]{JAC2003}

\usepackage{graphicx}
\usepackage{booktabs}

\usepackage{dcolumn, mathptmx}
\newcolumntype{.}{D{.}{.}{-1}} % centered on decimal point, including
\newcommand{\be}{\begin{equation}}
\newcommand{\ee}{\end{equation}}
\newcommand{\bea}{\begin{eqnarray}}
\newcommand{\eea}{\end{eqnarray}}

\newcommand{\q}[2]{\ensuremath{#1\ \mathrm{#2}}} % quantity with units
\newcommand{\dt}{\ensuremath{\Delta t}}
\newcommand{\PN}[1]{\ensuremath{P\left(\frac{#1}{\sigma}\right)}}
\newcommand{\e}[1]{\ensuremath{\exp{\left[-\frac{1}{2}\left(\frac{#1}{\sigma}\right)^2\right]}}}
\begin{document}
\title{BEAM HALO DYNAMICS AND\\ CONTROL WITH HOLLOW ELECTRON
  BEAMS\thanks{Fermilab is operated by Fermi Research Alliance, LLC
    under Contract No.~DE-AC02-07CH11359 with the United States
    Department of Energy.  This work was partially supported by the US
    LHC Accelerator Research Program (LARP).}}

\author{G.~Stancari\thanks{stancari@fnal.gov}, G.~Annala, A.~Didenko,
  T.~R.~Johnson, I.~A.~Morozov, V.~Previtali, G.~Saewert,\\
  V.~Shiltsev, D.~Still, A.~Valishev, L.~G.~Vorobiev,
  Fermilab, Batavia, IL 60510, USA\\
  D.~Shatilov, BINP, Novosibirsk, Russia\\
  R.~W.~Assmann\thanks{Present address: Deutsches
    Elektronen-Synchrotron (DESY), Hamburg, Germany.}, R.~Bruce,
  S.~Redaelli, A.~Rossi, B.~Salvachua~Ferrando, G.~Valentino,\\
  CERN, Geneva, Switzerland}

\maketitle

\begin{abstract}
  Experimental measurements of beam halo diffusion dynamics with
  collimator scans are reviewed. The concept of halo control with a
  hollow electron beam collimator, its demonstration at the Tevatron,
  and its possible applications at the LHC are discussed.
\end{abstract}

\section{INTRODUCTION}

Beam quality and machine performance in circular accelerators depend
on global quantities such as beam lifetimes, emittance growth rates,
dynamic apertures, and collimation efficiencies.  Calculations of
these quantities are routinely performed in the design stage of all
major accelerators, providing the foundation for the choice of
operational machine parameters.

At the microscopic level, the dynamics of particles in an accelerator
can be quite complex. Deviation from linear dynamics can be large,
especially in the beam halo. Lattice resonances and nonlinearities,
coupling, intrabeam and beam-gas scattering, and the beam-beam force
in colliders all contribute to the topology of the particles' phase
space, which in general includes regular areas with resonant islands
and chaotic regions. In addition, various noise sources are present in
a real machine, such as ground motion (resulting in orbit and tune
jitter) and ripple in the radiofrequency and magnet power supplies. As
a result, the macroscopic motion can acquire a stochastic character,
describable in terms of diffusion~\cite{Lichtenberg:1992,
  Chen:PRL:1992, Gerasimov:FERMILAB:1992, Zimmermann:PA:1994,
  Sen:PRL:1996}.

In this paper, we first address the issue of obtaining experimental
data on the dynamics of the beam halo. It was shown that beam halo
diffusion can be measured by observing the time evolution of particle
losses during a collimator scan~\cite{Seidel:1994}. These phenomena
were used to estimate the diffusion rate in the beam halo in the SPS
at CERN~\cite{Burnod:CERN:1990}, in HERA at DESY~\cite{Seidel:1994},
and in RHIC at BNL~\cite{Fliller:PAC:2003}. A much more extensive
experimental campaign was carried out at the Tevatron
in~2011~\cite{Stancari:IPAC:2011.diff} to characterize the beam
dynamics of colliding beams and to study the effects of the novel
hollow electron beam collimator
concept~\cite{Stancari:PRL:2011}. Recently, the technique was also
applied to measure halo diffusion rates in the LHC at
CERN~\cite{Stancari:FN:2012}. These measurements shed light on the
relationship between halo population and dynamics, emittance growth,
beam lifetime, and collimation efficiency. They are also important
inputs for collimator system design and upgrades, including new
methods such as channeling in bent crystals or hollow electron lenses.

In the second part of the paper, we discuss the novel concept of
hollow electron beam collimation (HEBC), and how it affects halo
dynamics. The results of experimental studies at the Fermilab Tevatron
collider are briefly reviewed, with an emphasis on the effect of the
hollow electron beam on halo diffusion in the circulating beam. We
conclude with a summary of recent research activities aimed at a
possible application of hollow beam collimation at CERN.

\section{BEAM HALO DIFFUSION}

As discussed in the introduction, particle motion in an accelerator at
the microscopic level is in general very rich. Two main considerations
lead to the hypothesis that macroscopic motion in a real machine,
especially in the halo, will be mostly stochastic: (1)~the
superposition of the multitude of dynamical effects (some of which
stochastic) acting on the beam; (2)~the operational experience during
collimator setup, which generates loss spikes and loss dips that often
decay in time as $1/\sqrt{t}$, a typically diffusive behavior.

\subsection{Experimental method}

\begin{figure}[b]
\begin{center}
\includegraphics[width=0.75\columnwidth]{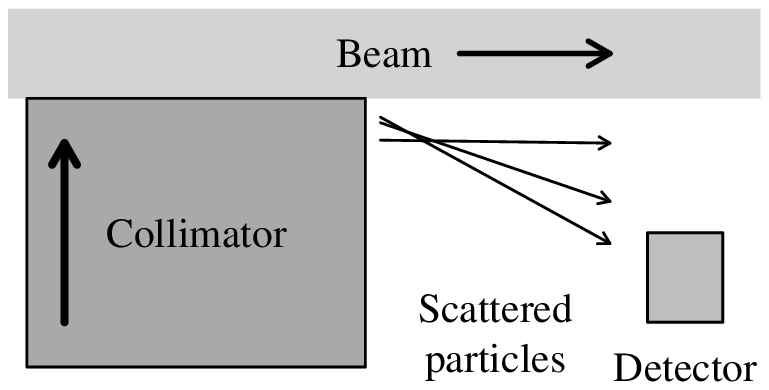}\\
\includegraphics[width=0.75\columnwidth]{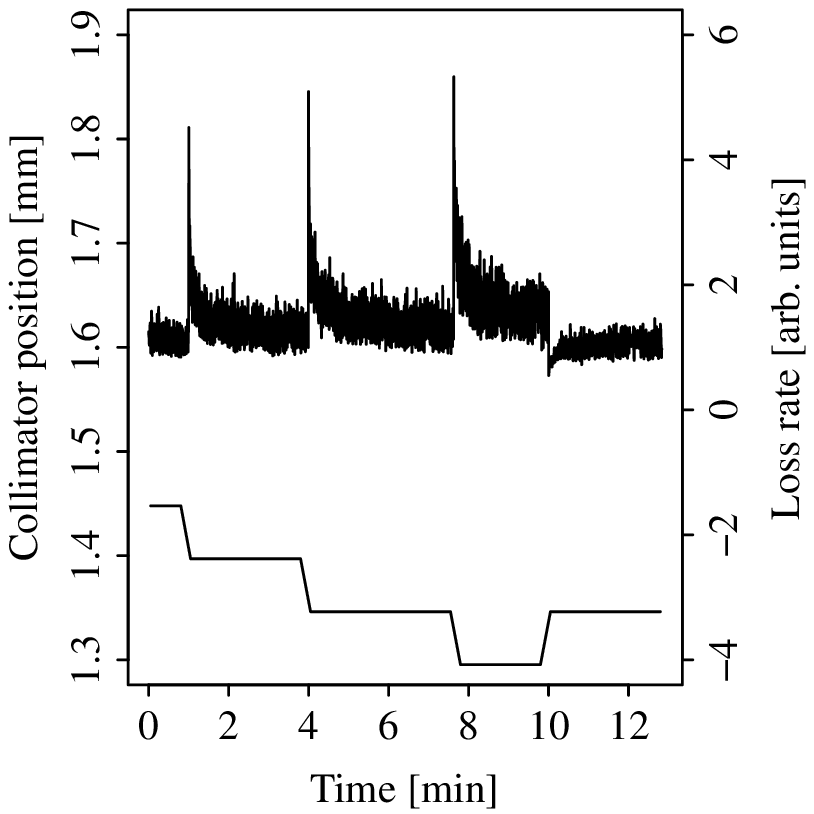}
\caption{Schematic diagram of the apparatus (top). Example of the
  response of local loss rates to inward and outward collimator steps
  (bottom).\label{fig:exp-method}}
\end{center}
\end{figure}

A schematic diagram of the apparatus is shown in
Fig.~\ref{fig:exp-method} (top).  All collimators except one are
retracted.  As the jaw of interest is moved in small steps (inward or
outward), the local shower rates are recorded as a function of
time. Collimator jaws define the machine aperture. If they are moved
towards the beam center in small steps, typical spikes in the local
shower rate are observed, which approach a new steady-state level with
a characteristic relaxation time (Fig.~\ref{fig:exp-method},
bottom). When collimators are retracted, on the other hand, a dip in
losses is observed, which also tends to a new equilibrium level.  By
using the diffusion model presented below, the time evolution of
losses can be related to the diffusion rate at the collimator
position. By independently calibrating the loss monitors against the
number of lost particles, halo populations and collimation
efficiencies can also be estimated.

\subsection{Model of loss rate evolution in a collimator scan}

A diffusion model of the time evolution of loss rates caused by a step
in collimator position was developed~\cite{Stancari:diff:2011}. It
builds upon the model of Ref.~\cite{Seidel:1994} and its assumptions:
(1)~constant diffusion rate and (2)~linear halo tails within the range
of the step. These hypotheses allow one to obtain analytical
expressions for the solutions of the diffusion equation and for the
corresponding loss rates vs.\ time. The present model addresses some
of the limitations of the previous model and expands it in the
following ways: (a)~losses before, during, and after the step are
predicted; (b)~different steady-state rates before and after are
explained; (c)~determination of the model parameters (diffusion
coefficient, tail population gradient, detector calibration, and
background rate) is more robust and precise.

Following Ref.~\cite{Seidel:1994}, we consider the evolution in
time~$t$ of a beam of particles with phase-space density~$f(J,t)$
described by the diffusion equation $\partial_t f = \partial_J \left(D
  \, \partial_J f \right)$, where~$J$ is the Hamiltonian action
and~$D$ the diffusion coefficient in action space. The particle flux
at a given location $J=J'$ is $\phi = -D \cdot \left[ \partial_J f
\right]_{J=J'}$.  During a collimator step, the action~$J_c = x^2_c /
(2 \beta_c)$, corresponding to the collimator half gap~$x_c$ at a ring
location where the amplitude function is~$\beta_c$, changes from its
initial value~$J_{ci}$ to its final value~$J_{cf}$ in a time~\dt. The
step in action is $\Delta J \equiv J_{cf} - J_{ci}$. In the Tevatron,
typical steps are \q{50}{\mu m} in 40~ms, and the amplitude function
is tens of meters.  It is assumed that the collimator steps are small
enough so that the diffusion coefficient can be treated as a constant
in that region. If~$D$ is constant, the local diffusion equation
becomes $\partial_t f = D \, \partial_{JJ} f$.  With these
definitions, the particle loss rate at the collimator is equal to the
flux at that location:
\be
L = -D \cdot \left[ \partial_J f \right]_{J=Jc}.
\label{eq:flux}
\ee
Particle showers caused by the loss of beam are measured with
scintillator counters or ionization chambers placed close to the
collimator jaw. The observed shower rate is parameterized as
\be
S = kL + B,
\label{eq:obs.rate}
\ee
where~$k$ is a calibration constant including detector acceptance and
efficiency and~$B$ is a background term which includes, for instance,
the effect of residual activation. Under the hypotheses described
above, the diffusion equation can be solved analytically using the
method of Green's functions, subject to the boundary condition of
vanishing density at the collimator and beyond. Details are given in
Ref.~\cite{Stancari:diff:2011}.

Local losses are proportional to the gradient of the distribution
function at the collimator. The gradients differ in the two cases of
inward and outward step, denoted by the~$I$ and~$O$ subscripts,
respectively:
\bea
\lefteqn{\partial_J f_I(J_c, t) = -A_i + 2(A_i - A_c) \PN{-J_c} +
\frac{2}{\sqrt{2\pi} \sigma} \cdot \mbox{}} \label{eq:gradI} \\
& & \left\{ -A_i(J_{ci}-J_c) +
 (A_i J_{ci}-A_c J_c) \e{J_c} \right\} \nonumber \\
\lefteqn{\partial_J f_O(J_c, t) = -2 A_i \PN{J_{ci}-J_c}
 + 2 (A_i - A_c) \PN{-J_c} + \mbox{}} \nonumber \\
& & \mbox{} + 2 \frac{A_i J_{ci} - A_c J_c}{\sqrt{2\pi} \sigma} \e{J_c}.
\label{eq:gradO}
\eea
The parameters~$A_i$ and $A_f$ are the slopes of the distribution
function before and after the step, whereas~$A_c$ varies linearly
between~$A_i$ and~$A_f$ as the collimator moves. The
parameter~$\sigma$ is defined as $\sigma \equiv \sqrt{2 D t}$. The
function~$P(x)$ is the S-shaped cumulative Gaussian distribution
function: $P(-\infty)=0$, $P(0)=1/2$, and $P(\infty)=1$.

The above expressions (Eqs.~\ref{eq:gradI} and~\ref{eq:gradO}) are
used to model the measured shower rates. Parameters are estimated from
a fit to the experimental data. The background~$B$ is measured before
and after the scan when the jaws are retracted. The calibration
factor~$k$ is in general a function of collimator position, and can be
determined independently by comparing the local loss rate with the
number of lost particles measured by the beam current transformer.
The parameters $(k D A_i)$ and $(k D A_f)$ depend on the steady-state
loss rate levels before and after the step. The diffusion
coefficient~$D$ is mainly influenced by the measured relaxation time
and by the value of the peak (or dip) in losses.

The model explains the data very well when the diffusion time is long
compared to the duration of the step. The model can be extended by
including a separate drift term (from the Fokker-Planck equation) or a
nonvanishing beam distribution at the collimator. With this technique
(collimator scans in small steps), the diffusion rate can be measured
over a wide range of amplitudes. At large amplitudes, the method is
limited by the vanishing beam population and by the fast diffusion
times. The limit at small amplitudes is given by the level of
tolerable loss spikes.

\subsection{Results}

\begin{figure}[b]
\begin{center}
\includegraphics[width=\columnwidth]{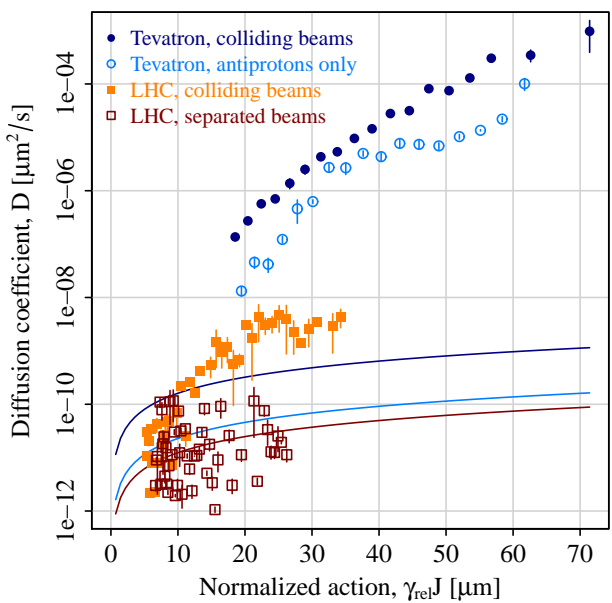}
\caption{Measurements of halo diffusion in the Tevatron and in the
  LHC.\label{fig:diff-meas}}
\end{center}
\end{figure}

Figure~\ref{fig:diff-meas} shows a comparison of beam halo diffusion
measurements in the Tevatron and in the LHC. These data sets refer to
vertical collimator scans. All Tevatron measurements were done on
antiprotons at the end of regular collider stores, either in
collisions (dark blue) or with only antiprotons in the machine (light
blue). The LHC measurements were taken in a special machine study at
4~TeV with only one bunch per beam, first with separated beams (red)
and then in collision (orange). To account for the different kinetic
energies, diffusion coefficients are plotted as a function of vertical
collimator action~$J$ multiplied by the relativistic Lorentz
factor~$\gamma_\mathrm{rel}$. The continuous lines represent the
diffusion coefficients calculated from the measured core geometrical
emittance growth rates~$\dot{\varepsilon}$: $D(J) = \dot{\varepsilon}
\cdot J$. (In this particular data set, the synchrotron-light
measurements were not sufficient to estimate emittance growth rates of
colliding beams in the LHC).

In the LHC, separated beams exibit a slow halo diffusion, comparable
with the emittance growth from the core.  This can be interpreted as a
confirmation of the extremely good quality of the magnetic fields in
the machine. Collisions enhance halo diffusion in the vertical plane
by about 1--2~orders of magnitude. In the Tevatron, the data suggests
that beam halo diffusion is dominated by effects other than
residual-gas scattering and beam-beam forces, pointing towards field
nonlinearities and noise.

From the measured diffusion coefficients, estimates of impact
parameters on the primary collimator jaws are
possible~\cite{Seidel:1994}.  One can also deduce the steady-state
density of the beam tails, with a procedure that is complementary to
the conventional static model based on counting the number of lost
particles at each collimator step.

\section{COLLIMATION WITH\\ HOLLOW ELECTRON BEAMS}

\begin{figure}[b]
\begin{center}
\includegraphics[width=\columnwidth]{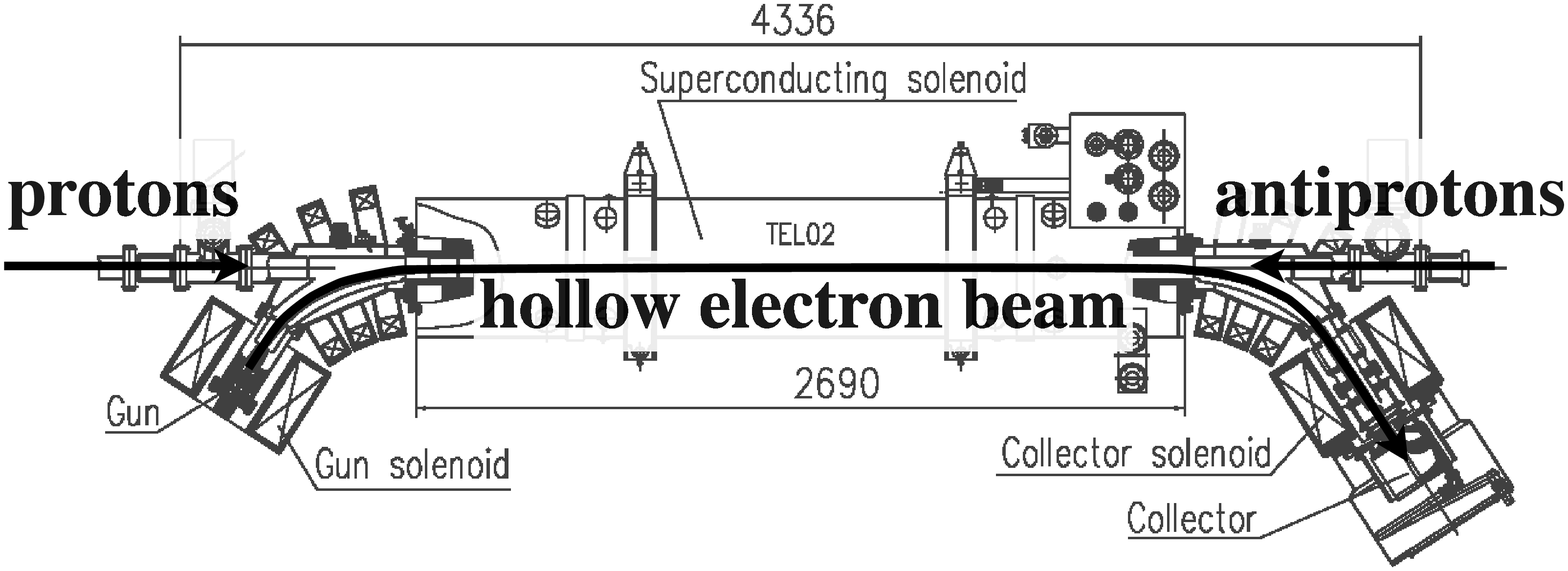}\\
\includegraphics[width=0.75\columnwidth]{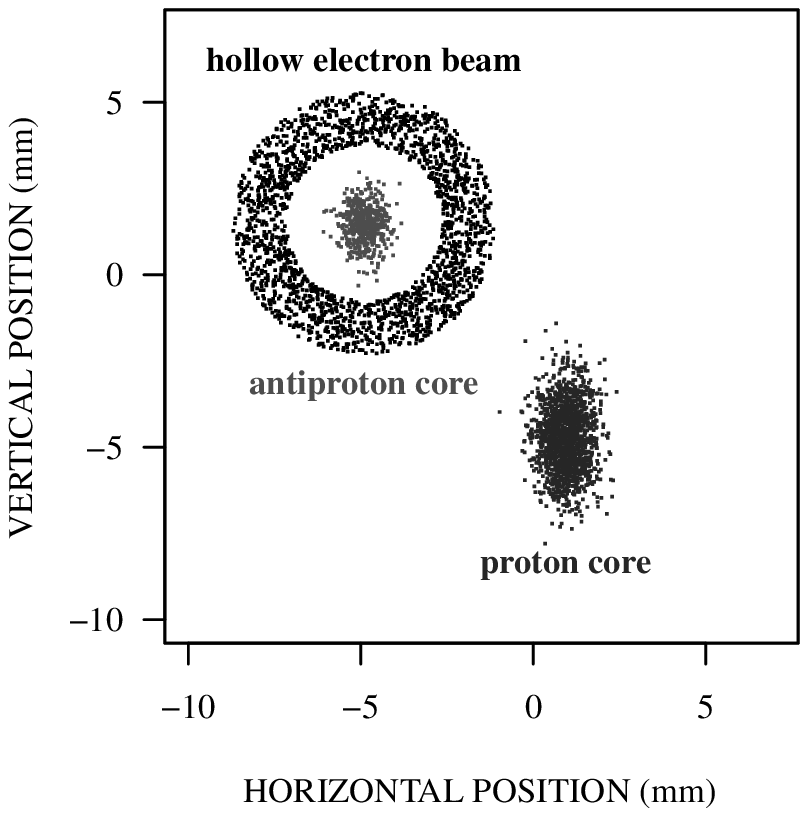}
\caption{Schematic diagram of the beam layout in the Tevatron hollow
  electron beam collimator.\label{fig:hebc}}
\end{center}
\end{figure}

In high-power accelerators, the stored beam energy can be large: about
2~MJ in the Tevatron, and several hundred megajoules in the LHC at
nominal energies and intensities. Uncontrolled losses of even a small
fraction of particles can damage components, cause magnets to lose
superconductivity, and increase experimental backgrounds. Contributing
to these losses is the beam halo, continuously replenished by beam-gas
and intrabeam scattering, ground motion, electrical noise in the
accelerating cavities, resonances and, in the case of colliders,
beam-beam forces. The beam collimation system is therefore vital for
the operation of high-power machines. Conventional collimation schemes
include scatterers and absorbers, possibly including several
stages. Primary collimators are the devices closest to the beam. They
impart random transverse kicks, mainly due to multiple Coulomb
scattering, to particles in the halo. The affected particles have
increasing oscillation amplitudes and a large fraction of them is
captured by the secondary collimators. These systems offer robust
shielding of sensitive components. They are also very efficient in
reducing beam losses at the experiments. A description
of the Tevatron and LHC collimation systems can be found in
Refs.~\cite{Mokhov:JINST:2011, Bruening:LHC:2004}.

The classic multi-stage systems have some limitiations: a fraction of
particles is always lost around the ring (leakage); collimator jaws
have an electromagnetic impedance (wakefields); and high losses are
generated during collimator setup when the jaws are moved inward to
scrape away the halo. Another problem is beam jitter. The orbit of the
circulating beam oscillates due ground motion and other
vibrations. This translates into periodic bursts of losses at aperture
restrictions. Hollow beams are a novel technique that addresses some
of these limitations.

\subsection{Concept}

The hollow electron beam collimator is a cylindrical, hollow,
magnetically confined, possibly pulsed electron beam overlapping with
the beam halo~\cite{Shiltsev:CERN:2007a, Shiltsev:CERN:2007b,
  Shiltsev:EPAC:2008, Stancari:PRL:2011}
(Fig.~\ref{fig:hebc}). Electrons enclose the circulating beam. Halo
particles are kicked transversely by the electromagnetic field of the
electrons. If the hollow charge distribution is axially symmetric, the
core of the circulating beam does not experience any electric or
magnetic fields. For typical parameters, the transverse kick given to
980-GeV protons or antiprotons in the Tevatron is of the order of
\q{0.2}{\mu rad}. This is to be compared with the multiple-scattering
random kick of \q{17}{\mu rad} from the primary tungsten collimators
in the Tevatron. With the hollow electron lens, one aims at enhancing
diffusion of the beam tails. This reduces their population in a
controllable way. It also decreases the loss spikes caused by
collimator setup, tune adjustments, and beam jitter.

A magnetically confined electron beam has several advantages. It can
be placed very close to, and even overlap with the circulating
beam. The transverse kicks are small and tunable, so that the device
acts more like a ``soft scraper'' or a ``diffusion enhancer,'' rather
than a hard aperture limitation.  At even higher electron currents
(which have not been demonstrated, yet) the electron beam could become
an indestructible primary collimator. If needed, the electron beam can
be pulsed to only affect a subset of bunches, or the abort gap (for
setup and alignment purposes, for instance, or for abort-gap
cleaning~\cite{Zhang:PRSTAB:2008}). If faster particle removal is
needed, the electron beam can be pulsed resonantly with the betatron
oscillations. In the case of ion collimation, there is no nuclear
breakup. Finally, the device relies on the estabilished technologies
of electron cooling~\cite{Parkhomchuk:RAST:2008} and electron
lenses~\cite{Shiltsev:PRL:2007, Shiltsev:NJP:2008,
  Shiltsev:PRSTAB:2008}. One disadvantage may be the relative cost and
complexity of the required components, such as superconducting
solenoids, high-voltage modulators, and cryogenics.

\subsection{Beam experiments at the Tevatron}

\begin{figure}[b]
\begin{center}
\includegraphics[width=\columnwidth]{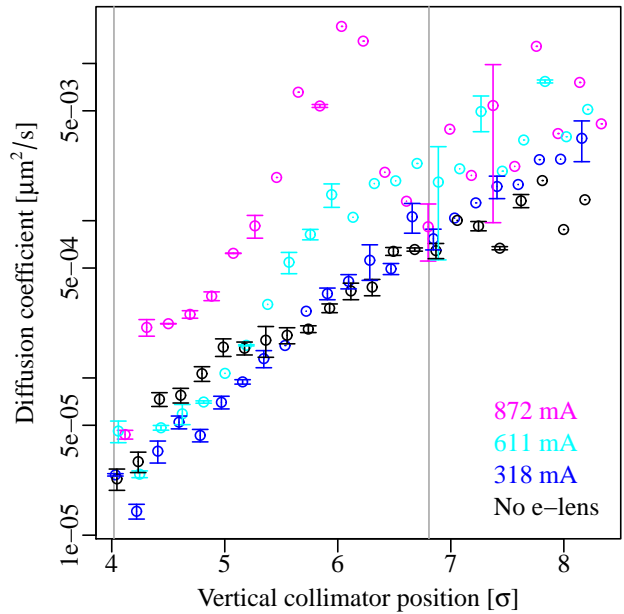}
\caption{Effect of the hollow electron beam on the transverse
  diffusion coefficient, as a function of vertical collimator
  position. The gray lines represent the calculated geometrical
  projection of the hollow electron beam.\label{fig:hebc-diff}}
\end{center}
\end{figure}

The concept of hollow electron beam collimation was tested in the
Fermilab Tevatron collider. Experiments in the Tevatron started in
October~2010 and ended with the shutdown of the machine in
September~2011. Many observables such as particle removal rates,
effects on the core, diffusion enhancement, collimation efficiency and
loss rate fluctuations were measured as a function of electron lens
parameters: beam current, relative alignment, hole radius, pulsing
pattern, and collimator configuration. Preliminary results were
presented in Refs.~\cite{Stancari:PRL:2011, Stancari:PAC:2011,
  Stancari:IPAC:2011.hebc, Stancari:APS:2011}. Here, we summarize the
main observations: (a)~compatibility with collider operations --- the
electron lens was routinely operated during regular collider stores
without loss of luminosity; (b)~reliable and reproducible alignment of
the electron beam with the circulating beam; (c)~smooth halo removal;
(d)~negligible effects on the core (no particle removal or emittance
growth); (e)~suppression of loss spikes due to beam jitter or tune
adjustments; (f)~increased collimation efficiency, defined as the
ratio between local collimator losses and losses at the experiments;
(g)~transverse diffusion enhancement.

In the context of this paper, we would like to focus on the
enhancement of beam halo diffusion. We are interested in how the
hollow beam affects diffusion. For this purpose, new scintillator
paddles were installed near one of the antiproton secondary
collimators. These loss monitors were gated to individual bunch
trains. With this device we could measure diffusion rates, collimation
efficiencies and loss spikes simultaneously for the bunch trains
affected by the electron beam and for the control bunch trains.

Figure~\ref{fig:hebc-diff} shows a measurement of the diffusion
coefficient using the collimator-scan technique described in the first
part of the paper. The measurement was taken at the end of a regular
collider store. The diffusion coefficient is plotted as a function of
the vertical collimator position (expressed in terms of the r.m.s.\
beam size~$\sigma$), for different values of the electron beam
current. One can see a clear diffusion enhancement (up to 2~orders of
magnitude for a beam current of 0.9~A) in the region of transverse
space where the electron beam is present.

\subsection{Applications to the LHC}

Following the Tevatron experience, hollow electron beams are being
considered as a complement to the LHC collimation system for operation
at high intensities. The main feature of this novel technique is a
safe and flexible control of beam tails. A first step towards this
goal could be the installation of one of the existing Tevatron
electron lenses in the SPS or in the LHC, where candidate locations
have been identified.

To elucidate the dependence of the electron lens effect on the details
of the machine, a campaign of numerical simulations is being carried
out~\cite{Morozov:IPAC:2012, Previtali:IPAC:2012}. The main goals are
to understand the Tevatron observations, to develop complementary
tools (Lifetrac and SixTrack), to check their consistency, and to
extend the simulations to test scenarios in the SPS or in the LHC.

It is also desirable to develop larger electron guns, for two reasons:
to achieve larger currents, and therefore extend the reach of the
hollow lens; and to operate at higher solenoidal fields, improving the
stability of the two-beam system. For these purposes, a new 1-inch
hollow electron gun was built and is being tested at Fermilab.

\section{CONCLUSIONS}

Collimator scans are a sensitive tool for the study of halo
dynamics. They allow one to investigate beam diffusion, populations,
lifetimes, and collimation efficiencies as a function of transverse
amplitude. Measurements of halo diffusion rates in the Tevatron and in
the LHC were presented, quantifying the role of the different
mechanisms dominating halo dynamics.

Hollow electron beams have been experimentally shown to be a safe and
flexible technique for halo control in high-power accelerators. In
this paper, their effect on halo dynamics in circular accelerators was
emphasized. The characteristics of each individual machine must be
taken into account to understand the details of their operation, but
the technique appears to be applicable to the LHC and other
accelerators.

\end{document}